\documentclass{article}
\usepackage{spconf,amsmath,graphicx,amssymb}
\usepackage{color,multirow}
\usepackage{cite}
\usepackage{type1cm}

\DeclareMathOperator\sign{sign}
\DeclareMathOperator\sg{sg}
\DeclareMathOperator\SDR{SDR}
\DeclareMathOperator\SIR{SIR}

\DeclareMathOperator\STPR{STPR}

\DeclareMathOperator\DIM{DI_{\mathcal{M}}}
\DeclareMathOperator\DSAM{DSA_{\mathcal{M}}}
\DeclareMathOperator\DSM{DS_{\mathcal{M}}}
\DeclareMathOperator\DISDR{DI_{SDR}}
\DeclareMathOperator\DSASDR{DSA_{SDR}}
\DeclareMathOperator\DSSDR{DS_{SDR}}
\DeclareMathOperator\DSSIR{DS_{SIR}}

\title{Adversarial attacks on audio source separation}
\name{
Naoya Takahashi$^1$, Shota Inoue$^{2*}$, Yuki Mitsufuji$^1$
}
\address{$^1$Sony Corporation, Japan  ~~~ $^2$University of Tsukuba, Japan }
\begin{document}
\ninept
\fontsize{9pt}{11.4pt}\selectfont


\maketitle
\begin{abstract}
Despite the excellent performance of neural-network-based audio source separation methods and their wide range of applications, their robustness against intentional attacks has been largely neglected. In this work, we reformulate various adversarial attack methods for the audio source separation problem and intensively investigate them under different attack conditions and target models. We further propose a simple yet effective regularization method to obtain imperceptible adversarial noise while maximizing the impact on separation quality with low computational complexity. Experimental results show that it is possible to largely degrade the separation quality by adding imperceptibly small noise when the noise is crafted for the target model. We also show the robustness of source separation models against a black-box attack. This study provides potentially useful insights for developing content protection methods against the abuse of separated signals and improving the separation performance and robustness. 
\end{abstract}
\begin{keywords}
audio source separation, adversarial example
\end{keywords}
\section{Introduction}
\label{sec:intro}

\begin{figure}[t]
  \centering
  \includegraphics[width=\linewidth]{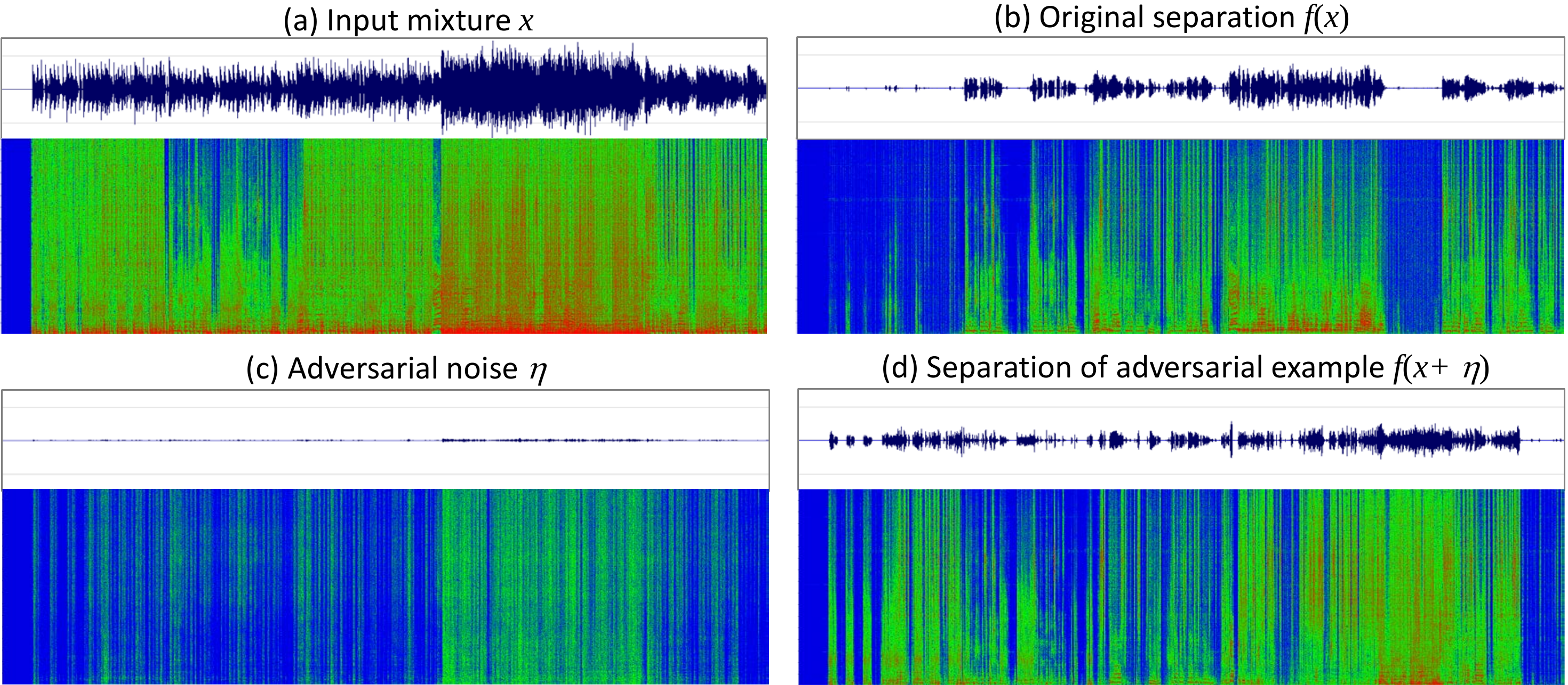}
  \caption{Visualization of adversarial noise and its effect in time-frequency domain. By adding the hardly perceptible adversarial noise (c) to the input (a), the separation degrades drastically (d) from the original separation (b). }
  \label{fig:visualize}
\end{figure}

\renewcommand{\thefootnote}{\fnsymbol{footnote}}
\footnote[0]{* Inoue contributed to the work while interning at Sony.}
Audio source separation has been intensively studied and widely used for downstream tasks. For instance, various music information retrieval tasks, including lyric recognition and alignment \cite{Mesaros2010,Fujihara2011, Sharma2019}, music transcription \cite{Gillet2008, Manilow2020}, instrument classification \cite{Gomez2018}, and singing voice generation \cite{Liu2019}, rely on music source separation (MSS). Likewise, automatic speech recognition benefits from speech enhancement and speech separation. 
Recent advances in source separation methods based on deep neural networks~(DNNs) have dramatically improved the accuracy of separation and some methods perform comparably to or even better than ideal-mask methods, which are used as theoretical upper baselines \cite{JanssonHMBKW17,Takahashi17,Takahashi18MMDenseLSTM,JLee2019,Liu2019mss,Luo18cTAS,Takahashi19,Takahashi20}. Although powerful DNN-based open-source libraries have become available \cite{Manilow2018,stoter19,defossez2019music, spleeter2020} and been used in the community, the robustness of source separation models against intentional attacks has been largely neglected. 
However, understanding the robustness against intentional attacks is important for the following reasons: (i) if one maliciously manipulates audio in perceptually undetectable ways such that the separation quality degrades severely, as shown in Fig.~\ref{fig:visualize}, all downstream tasks can fail; (ii) if creators do not want their audio contents to be separated and reused, such manipulation can protect contents from being separated with minimal and
imperceptible perturbation from the original content. The former is regarded as a defense against the attack on the separation model, while the latter as the copyright protection of content against the abuses of separated signals.   

In this work, we address this problem by investigating various adversarial attacks on audio source separation models. \textit{Adversarial examples} were originally discovered in an image classification problem where examples only slightly different from correctly classified examples drawn from the data distribution can be even confidently misclassified by a DNN \cite{Szegedy2014}. In \cite{Szegedy2014}, such adversarial examples were created by adding small perturbations called adversarial noise that maximize the image classification error. In many cases, such perturbations are hardly perceptible, and the adversarial examples generalize well across models with different architectures. As adversarial examples can be critical for many applications, they have been intensively investigated from different aspects including generation methods \cite{Su2019}, defending methods \cite{Madry2018, Bai2019}, transferability \cite{Dong2018, Wu2020}, and the cause of networks' vulnerabilities \cite{Goodfellow2015, Ilyas2019}.

Recently, adversarial attacks have also been investigated in audio domains including speech recognition \cite{Li2019, Schonherr2019}, speaker recognition \cite{Xie2020}, and audio event classification \cite{Subramanian2020}. However, all these works essentially address classification (or logistic regression) problems, and such models have similar properties, e.g., (i) they accept high-dimensional data such as a spectrogram or waveform and output a low-dimensional vector whose dimension is typically equal to the number of classes, (ii) their architecture typically employs a series of transformations from  high resolution with a few-channel representation to low resolution with many-channel representations, (iii) the class prediction is done through softmax.
In contrast, audio source separation is a regression problem and has very different properties: (i) the dimensionality of the output is high, typically the same as the input, (ii) the model may employ a transformation from a low-resolution representation to a high-resolution representation, (iii) the model does not necessarily incorporate softmax at the final output; the network can be trained to directly estimate real target values. Therefore, it is not clear if adversarial examples exist, how models behave against them, what type of attack is effective, and how much transferable the adversarial example is on the source separation problem. In this paper, we address these questions and intensively investigate a variety of attacks under different conditions. To our knowledge, this is the first work that investigates adversarial examples for the regression problem in the audio domain. 

The contributions of this work are summarized as follows:
(i) We reformulate adversarial attack methods for audio source separation; the reformulation does not require source signals to calculate the audio adversarial examples.
(ii) We propose a simple yet effective regularization method that can be used with reformulated attack methods by incorporating psychoacoustic masking effects.
(iii) We investigate the reformulated attacks by using well-known open-source MSS libraries and show how the source separation models are affected by the adversarial examples crafted by different attack methods under different conditions.  
(iv) We further investigate the transferability of adversarial examples to unseen models under black- and gray-box attack settings and to untargeted sources in a white-box setting.

\section{Methods: Adversarial Attacks against Source Separation Models}
\label{sec:attack}
\subsection{Gradient descent (GD)}
Adversarial examples were originally crafted by promoting the misclassification of image classification networks \cite{Szegedy2014}. In a similar way, we can define a perturbation \textbf{$\eta$} for an audio source separation network $f(\cdot)$ as a solution of a multidimensional regression problem:
\begin{equation}
    \max_{\eta \in \mathcal{D}} d(f(x+\eta),f(x)) ,~ \mathcal{D} = \{ \eta ~|~ \mathcal{C}(\eta) < \delta\},
    \label{eq:adv}
\end{equation}
where $x$ is the input audio, $\mathcal{C}$ is a constraint to limit the magnitude of the perturbation, and $\delta$ is a threshold value. A typical choice of $\mathcal{C}$ is the $l$2 norm $\|\eta\|_2$ or the supremum norm $\|\eta\|_\infty$. $d(\cdot,\cdot)$ is a metric and can be the $l1$ or $l2$ distance, or SI-SNR \cite{Isik16}. We used the $l2$ distance in this work. The motivation of Eq.~\eqref{eq:adv} is to craft a hardly perceptible perturbation $\eta$ that can maximize the difference of network outputs. It is worth noting that, unlike adversarial attacks in image or audio classification problems, Eq.~\eqref{eq:adv} does not require any label to estimate $\eta$, making the attack significantly practical because one can calculate an adversarial example without having access to the dataset on which the separation network $f$ is trained.  Eq.~\eqref{eq:adv} can be solved by minimizing the loss function $L$ by gradient descent:

\begin{equation}
    L(\eta) = -\|f(x+\eta)-f(x)\|_2^2 + \lambda \mathcal{C}(\eta),
    \label{eq:loss}
\end{equation}
where $\lambda$ is a scalar to control the regularization term. 

\subsection{Fast gradient sign method (FGSM)}
Goodfellow et~al. attempted to explain the cause of adversarial examples by hypothesizing the linear nature of a DNN \cite{Goodfellow2015}. Considering that the dot product of the perturbation $\eta$ and network weights $w$ can be maximized by assigning $\eta=\sign(w)$ under the max norm constraint, FGSM calculates the perturbation as
\begin{equation}
    \eta = \epsilon \sign(\nabla_x L(f(x), y)),
    \label{eq:fgsm}
\end{equation}
where $\epsilon$ is the magnitude of the perturbation, $L(a,b)$ is the loss function, and $y$ is a reference signal. In the original image classification setting  \cite{Goodfellow2015}, $L$ is the cross entropy loss and $y$ is the target class label. To apply FGSM for the audio source separation problem, we modify Eq.~\eqref{eq:fgsm} to use the mean square error loss $L(a,b)=\|a-b\|_2^2$ and provide the separated output with stopping gradient operation $\sg(f(x))$ as a reference signal.

\subsection{Projected gradient descent (PGD)}
Eq.~\eqref{eq:fgsm} can be seen as a single step scheme for maximizing the inner part of the saddle point formulation.
PGD extends and calculates the perturbation by $T$ iterative steps with smaller step size. After each perturbation step, PGD projects the adversarial example back onto the $\epsilon$-ball of $x$ if it goes beyond the $\epsilon$-ball. Similarly to FGSM, we apply PGD to the audio source separation problem as

\begin{equation}
    x^{t+1} = \Pi_\epsilon\left( x^t + \alpha \sign(\nabla_x L(f(x^t), \sg(f(x)))\right),
    \label{eq:pgd}
\end{equation}
where $\alpha$ is the step size and $\Pi_\epsilon$ denotes the projection operation to the $\epsilon$-ball.

\section{Conditions: Black- and gray-box attacks} 
\label{sec:bbattack}
The methods introduced in Sec.~\ref{sec:attack} assume that the gradient information of the target model is available for calculating the adversarial example. This setting is regarded as the white-box setting and the adversary has full access to the parameters of a target model.  While the white-box attack is applicable, for instance, on open-sourced software, compiled software usually does not give access to the gradient information. 
However, in some image classification problems, it is known that some adversarial examples crafted for a model are often effective for various models with different architectures or models trained on different subsets of training data \cite{Szegedy2014}. This property, called \textit{transferability}, is used to attack the target model without accessing the internal calculation pipeline (black-box setting). One can directly apply white-box methods to a surrogate model and use the created adversarial example to attack the target model.
Although the adversarial examples surprisingly generalize to untargeted models, the black-box attack is often less effective than the white-box attack.
One way to improve the transferability to the target model is to use prior knowledge about the target model architecture. In a gray-box setting, the network architecture of the target model is assumed to be known, while access to the network parameters is prohibited. 
Although the target gradient information is still unavailable in the gray-box setting, a model with the same architecture is assumed to provide more similar gradient to the target model than a model with a different architecture. Hence, the adversarial examples are expected to exhibit better transferability.

\section{Incorporating Psychoacoustic model} 
\label{sec:masking}
The perturbation $\eta$ is desired to be imperceptible. In image classification, this can be achieved by uniformly regularizing the magnitude of perturbation in terms of the $l$2 norm $\|\eta\|_2$ or the supremum norm $\|\eta\|_\infty$. However, in audio source separation, this is not the optimal choice since the perceptibility of the perturbation depends highly on the input signals. For example, low-level noise can be highly perceptible in silent regions, while high-level noise can be hardly audible when a high-level signal exists. This phenomenon is referred to as the \textit{masking effect}, where a louder signal can make other signals at nearby frequencies (\textit{frequency masking}) or time (\textit{time masking}) imperceptible. Previous works attempted to incorporate the masking effect by using an external MP3 encoder \cite{Xie2020} or by the iterative estimation of masking thresholds \cite{Qin2019}. However, the optimization of a loss function with such a regularization term is often difficult and slow to converge. Here, we propose a simple yet effective regularization method using the short-term power ratio (STPR) of the input and adversarial noise as
\begin{equation}
   \mathcal{C}_{\STPR}(\eta) = \|\vartheta(\eta, l) / \vartheta(x, l)\|_1,
    \label{eq:mask}
\end{equation}
where $\vartheta(\eta, l) = [\bar{\eta}_1, \cdots, \bar{\eta}_N]$ is the patch-wise $l2$ norm function with window length $l$,  $\bar{\eta}_n=\|(\eta_{(n-1)l},\cdots,\eta_{nl})\|_2$ is the $l2$ norm of patch index $n$, and $\eta_t$ is the sample value at time $t$.
$\vartheta$ can be implemented by using an average pooling function, which is commonly available in deep learning libraries. In contrast to previous methods, $\mathcal{C}_{\STPR}(\eta)$ does not involve an external MP3 encoder or an iterative estimation process of the masking threshold and can be efficiently calculated, which is particularly important for iterative methods such as GD and PGD.   Although Eq.~\eqref{eq:mask} does not explicitly consider the masking threshold, we found that it regularizes the adversarial signal well enough to achieve hardly perceptible adversarial noise depending on the input signal level without hindering the convergence. 


\section{Experiments}
\label{sec:exSS}
\subsection{Setup}
We conducted experiments on the \textit{test} set of MUSDB18 dataset \cite{sisec2018}. 
In the dataset, a mixture and its four sources, {\it bass, drums, other}, and {\it vocals}, recorded in stereo format at 44.1 kHz, are available for each song. We used \textit{vocals} as a target instrument for attacks.   
Two open-source MSS libraries are considered, namely, Open-Unmix (UMX)\cite{stoter19} and Demucs \cite{defossez2019music}. UMX performs separation in the frequency domain with bidirectional LSTM layers, while Demucs performs separation in the time domain with convolutional neural networks (CNNs). 
We used publicly available pretrained models.

\subsection{Evaluation metric}
Two important factors for evaluating adversarial examples are (i) \textit{How much does the adversarial example degrade the separation quality?} and (ii) \textit{How strong (or perceptible) is the perturbation?}  
To objectively evaluate these factors, we define three metrics: degradation of separation $\DSM$, degradation of input $\DIM$, and degradation of separation with additive adversarial noise $\DSAM$. Degradation is measured based on a \textit{ground metric} $\mathcal{M}$, where $\mathcal{M}$ can be the signal-to-distortion ratio (SDR), signal-to-interference ratio (SIR), or other metrics.  For clarity, we consider in the following $\mathcal{M}=\SDR$. Let $x,y$, and $\eta$ be the input mixture, target source, and adversarial noise, respectively, and let $\SDR(y,\hat{y})$ be the SDR of $\hat{y}$ with reference $y$. 
We define
\begin{eqnarray}
    \DSSDR &=& \SDR(y, f(x)) - \SDR(y, f(x+\eta)),\\
    \DISDR &=& \SDR(x, x+\eta), \\
    \DSASDR &=& \SDR(y, f(x)) - \SDR(y, f(x)+\eta).
    \label{eq:metric}
\end{eqnarray}
$\DSSDR$ indicates how much the SDR is degraded by the adversarial example $x+\eta$ compared with the separation of the original mixture $x$. Higher $\DSSDR$ means that the adversarial example degrades the separation more significantly.  $\DISDR$ and $\DSASDR$ are the evaluation metrics for the noise level. $\DISDR$ directly evaluates the SDR of the adversarial example against the original input while $\DSASDR$ evaluates how much the SDR is degraded if the adversarial noise is directly added to the original separation $f(x)$. Similarly, we define, e.g., $\DSSIR$ using $\mathcal{M}=\SIR$. 
SDR and SIR values were computed using the {\it museval} package and the median over all tracks of the median of the metric over each track is reported, as in SiSEC 2018 \cite{sisec2018}.

\subsection{White-box attack}
\begin{figure}[t]
  \centering
  \includegraphics[width=\linewidth]{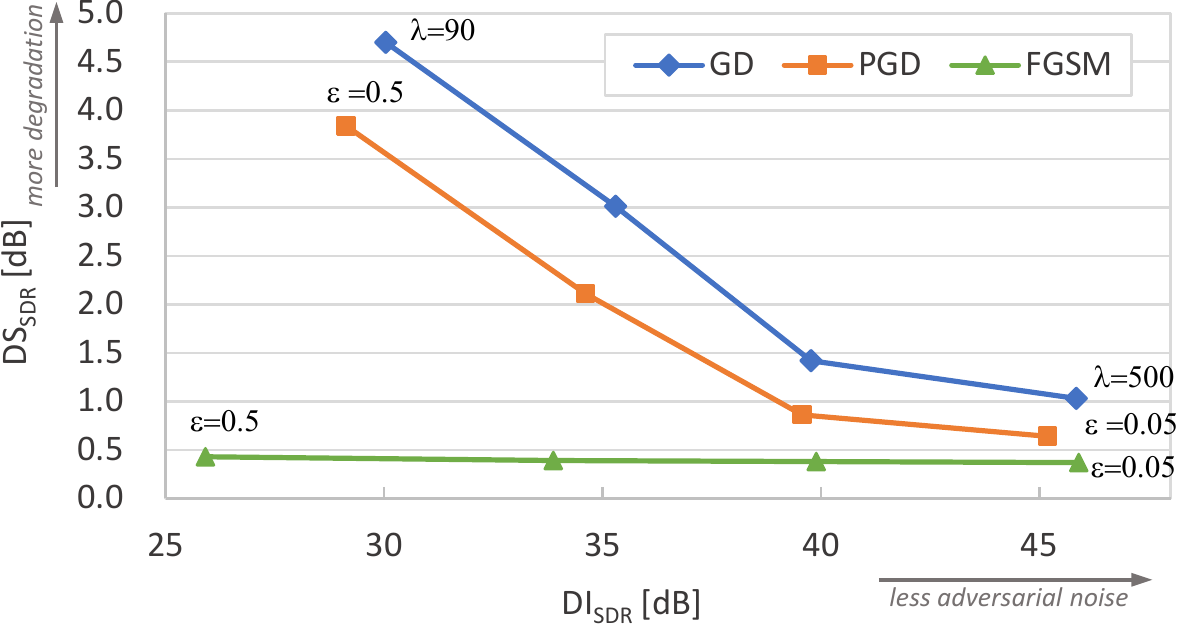}
  \caption{Comparison of attack methods against UMX with different adversarial noise levels.}
  \label{fig:attacks}
\end{figure}
\textbf{Attack method.} \hspace{1mm}
First, we compared three attack methods, namely, GD, PGD, and FGSM, with different adversarial noise levels. For GD, we performed 300 iterations and $\lambda$ was set to $(90, 170, 290,500)$.
For PGD and FGSM, $\epsilon$ was set to (0.05, 0.1, 0.2, 0.5). $\alpha$ for PGD was set to $\epsilon/\sqrt{k}$, where $k$ is the number of iterations. Results of attacks against UMX are shown in Fig. \ref{fig:attacks}. The SDR degradation becomes more prominent as the adversarial noise level becomes high for GD and PGD, while the $\DSSDR$ of FGSM remains around 0.4 dB. 
This suggests that FGSM is not very effective in attacking the source separation model. 
Both GD and PGD introduce significant SDR degradation with very low level adversarial noise. 
Since GD consistently led to more significant SDR degradation in the entire input SDR range than PGD, we used GD for the rest of our experiments.
\vspace{3mm}\\
\textbf{Power ratio regularization.} \hspace{1mm}
Fig.~\ref{fig:visualize} shows an example of the waveform and the spectrogram of (a) the input mixture, (b) the original separation, (c) the adversarial noise, and (d) the separation of the adversarial example. The target model was Demucs. By comparing the mixture with the adversarial noise, it can be seen that noise does not exist in the silent region at the beginning but becomes more prominent when the level of the input mixture is high. This helps to make the adversarial noise imperceptible by the masking effect. When we regularize the adversarial noise with the $l2$ norm, we obtain adversarial noise that spreads across time and is much more audible in silent or low-level input regions than adversarial noise crafted with the proposed STPR regularization with a similar $\DSSDR$. 
To validate this, we conducted a subjective test in similarly to the double-blind triple-stimulus with hidden reference format (ITU-R BS.1116), where the reference was the original mixture and either A or B was the same as the reference and the other was an adversarial example crafted with either the $l2$ or STPR regularization. The subjects were asked to identify which of A and B was the same as the reference signal. Twelve participants evaluated nine audio clips of 6 s duration, resulted in $99$ evaluations for each method. The results in Table~\ref{tab:subj} show that the accuracy of identifying the adversarial examples crafted with the proposed STPR is almost equal to the chance rate (50\%), showing that the STPR successfully produced inaudible adversarial examples, while the adversarial examples crafted with $l2$ regularization that have almost the same $\DSSDR$ as the STPR can be identified frequently.
This shows that the simple STPR is good enough to obtain hardly perceptible adversarial noise that significantly degrades the separation quality by considering the masking effect. 
\begin{table}[t]
    \caption{\label{tab:subj} {\it Comparison of regularization methods.}}
    \vspace{2mm}
    \centering{
    \begin{tabular}{c | c c | c} 
    \hline
    Method & $\DISDR$ [dB] & $\DSSDR$  [dB] &	Accuracy\\
    \hline\hline
    l2	& 28.83 &	5.25 &	75.9\%\\
    STPR &	30.33 &	5.23 &	52.8\%\\
    \hline
    \end{tabular}
    }
\end{table}
\vspace{3mm}\\
\textbf{Target models and attack domain.} \hspace{1mm}
We also compared the time domain model (Demucs) and frequency domain model (UMX). For UMX, we computed the adversarial example either in the time domain by back-propagating the error through the short-time Fourier transform (STFT) operation or in the frequency domain and transformed the obtained adversarial example to the time domain signal using the Griffin--Lim algorithm. Table \ref{tab:domain} shows that for all the target models and noise injection domains, subtle noise with $\DSASDR$ of less than 0.05 dB  significantly degraded the SDRs. 
Demucs is much more prone to adversarial attacks than UMX. 
Comparing the results for the adversarial noise calculation domain for attacking UMX, designing in the frequency domain is slightly more effective as it achieved higher $\DSSDR$ with higher $\DISDR$.
\begin{table}[t]
    \caption{\label{tab:domain} {\it Comparison of model and domain difference (in dB).}}
    \vspace{2mm}
    \centering{
    \begin{tabular}{c c | c c c c} 
    \hline
    Model & domain &	$\DISDR$ &	$\DSSDR$  &	$\DSSIR$& $\DSASDR$\\
    \hline\hline
    UMX &	freq. &	37.04 &	2.66 & 3.72 &	0.01\\
    UMX	& time &	36.70 &	2.50 & 4.14 &	0.04\\
    Demucs &	time &	37.08 &	5.83 & 14.13 &	0.03\\
    \hline
    \end{tabular}
    }
\end{table}
\vspace{3mm}\\
\textbf{Effects on untargeted instruments.} \hspace{1mm}
UMX and Demucs are trained to separate the four sources as defined in MUSDB. Therefore, it is interesting to see how the adversarial example crafted for one target instrument affects the separation of untargeted instruments. For this, we compared $\DSSDR$ values of the four instruments in Table \ref{tab:instr}. As observed, the effects on untargeted instruments depend on the instrument, e.g., \textit{bass} has only negligible effects while \textit{other} exhibits some degradation. This implies that the instruments whose frequency characteristics overlap with the target instrument could be more impacted than non-overlapping instruments. 
\begin{table}[t]
    \caption{\label{tab:instr} {\it $\DSSDR$ comparison of target (vocals) and untargeted instruments (in dB).}}
    \vspace{2mm}
    \centering{
    \begin{tabular}{c | c | c c c} 
    \hline
    Model & vocals & drums & bass & other\\
    \hline\hline
UMX	& 2.66 &	0.14	& 0.06 &	0.67\\
Demucs &	5.83 &	1.30 &	0.09 &	3.19\\
    \hline
    \end{tabular}
    }
\end{table}	
\vspace{3mm}\\
\textbf{Discussion.} \hspace{1mm}
We observed that the separation of the adversarial example results in degradation in which the target source is suppressed or the level of contamination of other instrument sounds in the mixture is increased, but not degradation involving the creation of irrelevant artificial noise. 
This is interesting because Eq.~\eqref{eq:loss} does not impose any such constraint and it can be maximized by, for example, introducing a stationary noise. 
The results could be reasonable for mask-based separation approaches, where masks are estimated by separation models and applied to the input mixture to obtain the separated signal. The mask-based approaches include frequency-domain methods that use Wiener filtering (WF) such as UMX or time-domain approaches such as Conv-TasNet\cite{Luo18cTAS}. However, we observed the same tendency for non-mask-based approaches, which directly estimate the target signal in the time or frequency domain without explicit masking scheme, such as Demucs or MDenseNet \cite{Takahashi18} without WF. We hypothesize that even for the non-mask-based approaches, the networks learn the masking strategy internally, and therefore the adversarial examples are crafted in a way that increases the interference of other sources in the mixture or suppresses the target source. 

\subsection{Black- and gray-box attacks}
Finally, we investigated the robustness of audio source separation models against black- and gray-box attacks. To this end, the adversarial examples were first crafted for the source models and evaluated on target models that have different network architecture or parameters. In addition to UMX and Demucs, we also used TASNet and Demucs$_{ex}$, available in \cite{defossez2019music}, as target models. TASNet is another time-domain MSS model trained on MUSDB, while Demucs$_{ex}$ has same the network architecture as Demucs but was trained with extra data, and thus Demucs$_{ex}$ was used for the evaluation of the gray-box attack. Table \ref{tab:blackbox} shows the $\DSSDR$ values under the condition of $\DISDR\simeq37$~dB. 
The results show that the $\DSSDR$ of the black-box attack is much lower than that of the white-box attack. This suggests the robustness of source separation models against black-box attack, or in other words, adversarial examples are less transferable, which indicates robustness of systems that involve source separation. In contrast, to protect the audio content from the abuse of the separation, the transferability of adversarial examples should be be improved. 
Comparing the target models TASNet and UMX with the source model Demucs, TASNet had stronger impact than UMX, probably owing to the similarity of their network architectures since both Demucs and TASNet work on time-domain signals and their architectures are based on a CNN. This claim is further supported by the results of the gray-box attack as Demucs$_{ex}$ had stronger impact than the black-box attack. 

\begin{table}[t]
    \caption{\label{tab:blackbox} {\it Comparison of white-, gray-, and black-box attacks.}}
    \vspace{2mm}
    \centering{
    \begin{tabular}{c c c | c c} 
    \hline
    Condition & Source & Target	& $\DSSDR$~[dB] & $\DISDR$~[dB]\\
    \hline\hline
    \multirow{3}{*}{black-box} &	UMX & Demucs & 0.33 & 37.04\\
         & Demucs & UMX &	0.19 &	37.08\\
         & Demucs & TASNet & 0.52 &	37.08\\
    \hline
    gray-box &	Demucs & Demucs$_{ex}$ & 1.20 &	37.08\\
    \hline
    \multirow{2}{*}{white-box} &	Demucs & Demucs & 5.83 & 37.08\\
        & UMX & UMX & 2.66 & 37.04\\
    \hline
    \end{tabular}
    }
\end{table}

\section{Conclusion}
We investigated various adversarial attacks under various conditions on audio source separation methods by reformulating the adversarial attack methods on classification models. To achieve imperceptible adversarial noise while maximizing the impact with low complexity, we proposed a simple short-term power ratio regularization. Extensive experimental results show that some adversarial attack methods can significantly degrade the separation performance with imperceptible adversarial noise under the white-box condition, while source separation models exhibit robustness under the black-box condition. 

\ninept

\bibliographystyle{IEEEbib}
\bibliography{MIR,bss,other,advexamp}


\end{document}